\documentclass[12pt]{article}
\usepackage{mathrsfs}
\usepackage{natbib,graphicx,setspace,lscape,longtable,xr}
\usepackage{mathrsfs,amsmath,amsthm,amssymb,color}
\usepackage{natbib,epsfig,graphicx,pdfpages}
\usepackage{rotating}
\usepackage{float}
\usepackage{bm}
\usepackage{CJK}
\usepackage{ragged2e}
\usepackage{setspace}
\usepackage{threeparttable}
\usepackage{multirow}
\usepackage[shortlabels]{enumitem}
\usepackage{etoolbox}
\usepackage{subfig}
\usepackage[ruled]{algorithm2e}
\usepackage{adjustbox}
\usepackage{blindtext}
\usepackage{xcolor}
\usepackage{hyperref}
\hypersetup{
	linkcolor  = blue,
	citecolor  =blue,
	urlcolor   = blue,
	colorlinks = true,
}

\bibpunct{(}{)}{;}{a}{,}{,}

\newtheorem{theorem}{Theorem}
\newtheorem{lemma}{Lemma}
\newtheorem{proposition}{Proposition}
\newcommand{\mylabel}[2]{#2\def\@currentlabel{#2}\label{#1}}

\def\beq{\begin{equation}}
	\def\eeq{\end{equation}}
\def\beqr{\begin{eqnarray}}
	\def\eeqr{\end{eqnarray}}
\def\beqrs{\begin{eqnarray*}}
	\def\eeqrs{\end{eqnarray*}}
\def\bet{\begin{theorem}}
	\def\eet{\end{theorem}}
\def\bel{\begin{lemma}}
	\def\eel{\end{lemma}}
\def\bep{\begin{proposition}}
	\def\eep{\end{proposition}}
\def\bg{\begin{figure}[tbph]\begin{center}}
		\def\eg{\end{center}\end{figure}}

\def\bc{\begin{center}}
	\def\ec{\end{center}}

\DeclareMathOperator*{\argmin}{arg\,min}

\def\1{\mbox{\boldmath $1$}}
\def\0{\boldsymbol{0}}

\def\E{\mathbb E}

\def\E{\mathbb E}

\def\R{\mathbb R}

\usepackage[a4paper,bindingoffset=0.2in,%
left=1in,right=1in,top=1in,bottom=1in,%
footskip=.25in]{geometry}

\newcommand{\bds}{\boldsymbol}

\def\boxit#1{\vbox{\hrule\hbox{\vrule\kern6pt\vbox{\kern6pt#1\kern6pt}\kern6pt\vrule}\hrule}}

\numberwithin{equation}{section}

\begin{document}
	
	\begin{center}
		{\bf\Large A review of distributed statistical inference}\\
		Yuan Gao$^{a}$, Weidong Liu$^{b}$, Hansheng Wang$^{c}$, Xiaozhou Wang$^{a}$\\Yibo Yan$^{a}$,  Riquan Zhang$^{a}$\footnote{Corresponding author.\ E-mail address: rqzhang@stat.ecnu.edu.cn}\\
		{\it \footnotesize
		$^{a}$School of Statistics and Key Laboratory of Advanced Theory and Application in Statistics and Data Science - MOE, East China Normal University, Shanghai, China; 
		$^{b}$School of Mathematical Sciences and Key Lab of Artiﬁcial Intelligence - MOE, Shanghai Jiao Tong University, Shanghai, China; 
		$^{c}$Guanghua School of Management, Peking University, Beijing, China 
	}

	\end{center}

\begin{abstract}
	The rapid emergence of massive datasets in various fields poses a serious challenge to traditional statistical methods. Meanwhile, it provides opportunities for researchers to develop novel algorithms. Inspired by the idea of divide-and-conquer, various distributed frameworks for statistical estimation and inference have been proposed. They were developed to deal with large-scale statistical optimization problems. This paper aims to provide a comprehensive review for related literature. It includes parametric models, nonparametric models, and other frequently used models. Their key ideas and theoretical properties are summarized. The trade-off between communication cost and estimate precision together with other concerns are discussed.
\end{abstract}


	\newpage
	
	
\section{Introduction}

With the rapid development of information technology, datasets of massive sizes become increasingly available. E-commerce companies like Amazon have to analyze billions of transaction data for personalized recommendation. Bioinformatics scientists need to locate relevant genes corresponding to some specific phenotype or disease from massive SNPs data. For Internet related companies, large amounts of text, image, voice, and even video data are in urgent need of effective analysis. Due to the accelerated growth of data sizes, the computing power and memory of one single computer are no longer sufficient. Constraint on network bandwidth and other privacy or security considerations also make it difficult to process the whole data on one central machine. Accordingly, distributed computing systems become increasingly popular.

Similar to parallel computing executed on a single computer, distributed computing is closely related to the idea of divide-and-conquer. Simply speaking, for some statistical problems, we can divide a complicated large task into many small pieces so that they can be tackled simultaneously on multiple CPUs or machines. Their outcomes are then aggregated to obtain the final result. It is conceivable that this procedure can save the computing time substantially if the algorithm can be executed in a parallel way. The main difference between a traditional parallel computing system and a distributed computing system is the way they access memory. For parallel computing, different processors can share the same memory. Consequently, they can exchange information with each other in a super-efficient way. While for distributed computing, distinct machines are physically separated. They are often connected by a network. Accordingly, each machine can only access its own memory directly. Therefore, the inter-machine communication cost in terms of time spending could be significant and thus should be prudently considered.

The rest of this article is organized as follows. Section 2 studies parametric models. Section 3 focuses on nonparametric methods. Section 4 expresses some other related methods. The article is concluded with a short discussion in Section 5.

\section{Parametric models}

Assume a total of $N$ observations denoted as $Z_i = (X_i^\top,Y_i)^\top  \in \R^{p+1}$ with $ 1\le i\le N $. Here $X_i\in\R^p$ is the covariate vector and $Y_i\in\R$ is the corresponding scalar response. Define $\{\mathbb{P}_{\bds{\theta}}: \bds{\theta} \in \Theta \}$ to be a family of statistical models parameterized by $ \bds{\theta}\in \Theta \subset\R^p $. We further assume that $Z_i$'s are independent and identically distributed with the distribution $\mathbb{P}_{\bds{\theta}^*}$, where $ \bds{\theta}^*=(\theta_1^*,\dots,\theta_p^*)^\top $ is the true parameter. Consider a distributed setting, where $N$ sample units are allocated randomly and evenly to $K$ local machines  $\mathcal{M}_k,\ 1\le k \le K$, such that each machine has $n$ observations. Obviously, we should have $ N = nK $. Write $\mathbb{S}=\{1,\dots, N\}$ as the index set of whole sample. Then, let $\mathcal{S}_k$ denote the index set of local sample on $\mathcal{M}_k$ with $\mathcal{S}_{k_1}\cap\mathcal{S}_{k_2}=\emptyset$ for any $k_1\ne k_2$. Other than the local machines, there also exists a central machine represented by $ \mathcal{M}_\text{center} $. A standard architecture should have $ \mathcal{M}_\text{center} $ to be connected with every $ \mathcal{M}_k $.

Let $\mathcal{L}:\Theta\times\R^{p+1}\mapsto\R$ be the loss function. Assume that the true parameter $ \bds{\theta}^* $ minimizes the population risk $\mathcal{L}^*(\bds{\theta})=\E[\mathcal{L}(\bds{\theta};Z)]$,
where $ \E $ stands for expectation with respect to $ \mathbb{P}_{\bds{\theta}^*} $. Define the local loss on the $ k $th machine as $\mathcal{L}_{k} (\bds{\theta}) = n^{-1}\sum_{i\in\mathcal{S}_k} \mathcal{L}(\bds{\theta} ; Z_i)$. Correspondingly, define the global loss function based on the whole sample as $ \mathcal{L}(\bds{\theta})=N^{-1}\sum_{i\in\mathbb{S}} \mathcal{L}(\bds{\theta};Z_i)=K^{-1}\sum_{k=1}^{K}\mathcal{L}_k(\bds{\theta})$, whose minimizer is $ \hat{\bds{\theta}} = \argmin_{\bds{\theta}\in\Theta} \mathcal{L}(\bds{\theta})$.
In most cases, the whole sample estimator $ \hat{\bds{\theta}} $ should be $ \sqrt{N} $-consistent and asymptotically normal \citep{lehmann2006theory}. If $ N $ is small enough so that the whole sample $ \mathbb{S} $ can be read into the memory of one single computer, then $ \hat{\bds{\theta}} $ can be easily computed. The entire computation can be executed in the memory of this computer. On the other hand, if $ N $ is too large so that the whole sample $ \mathbb{S} $ cannot be placed on any single computer, then a distributed system must be used. In this case, the whole sample estimator $ \hat{\bds{\theta}} $ is no longer computable (or at least very difficult to compute) in practice. Then, how to develop novel statistical methods for distributed systems becomes a problem of great importance.

\subsection{One-shot approach}
To solve the problems, various methods have been proposed. They can be roughly divided into two classes. The first class contains so-called one-shot methods. They are to be reviewed in this subsection. The other class contains various iterative methods. They are to be reviewed in the next subsection.

\begin{figure}[h]
	\centering
	\subfloat[Illustration of the one-shot approach.]{%
		\resizebox*{0.45\textwidth}{!}{\includegraphics{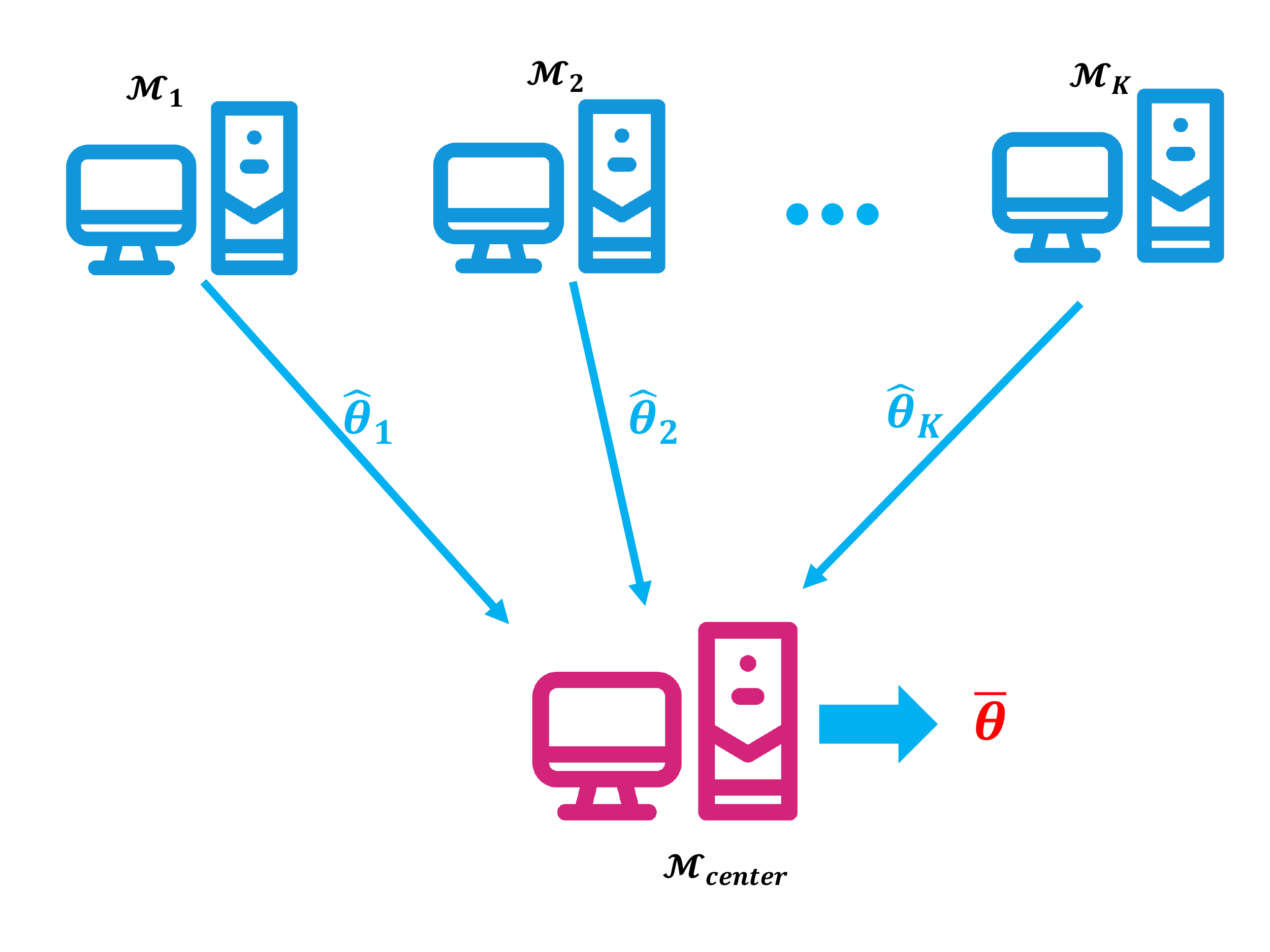}}}\hspace{5pt}
	\subfloat[Illustration of the iterative approach.]{%
		\resizebox*{0.45\textwidth}{!}{\includegraphics{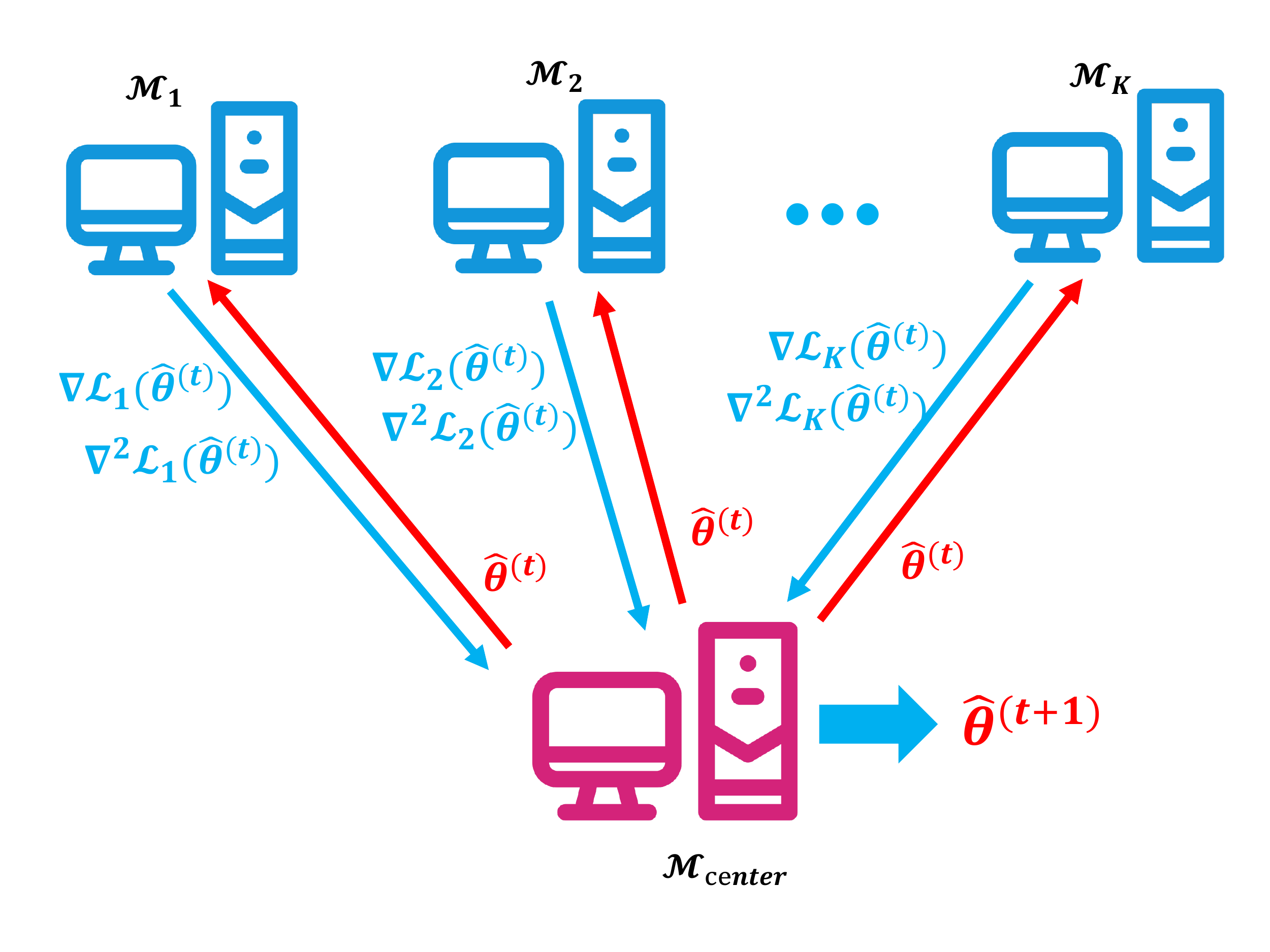}}}
	\caption{Illustrations of the two different approaches.} \label{fig:oneshot and iteative}
\end{figure}

The basic idea of the one-shot approach is to calculate some relevant statistics on each local machine. Subsequently, they are sent to a central machine, where these statistics are assembled into the final estimator. The most popular and direct way of aggregation is simple average. Specifically, for each $1\le k \le K$, machine $\mathcal{M}_k$ uses local sample $ \mathcal{S}_k $ to compute the local empirical minimizer $ \hat{\bds{\theta}}_k = \argmin_{\bds{\theta}\in\Theta} \mathcal{L}_k(\bds{\theta})$. These local estimates (i.e., $ \hat{\bds{\theta}}_k$'s) are then transferred to the center machine $\mathcal{M}_{\text{center}}$, where they are averaged as $ \bar{\bds{\theta}}=K^{-1}\sum_{k=1}^K \hat{\bds{\theta}}_k $. This leads to the final simple averaging estimator $ \bar{\bds{\theta}}$ (see Figure \ref{fig:oneshot and iteative}(a)).

Obviously, the one-shot style of distributed framework is highly communication-efficient. Because it requires only one single round of communication between each $ \mathcal{M}_k $ and $\mathcal{M}_{\text{center}}$. Hence, the communication cost is of the order $ O(Kp) $, where $ p $ is the dimension of each estimate $ \hat{\bds{\theta}} $. Theoretical properties of simple averaging estimator were also studied in the literature.  For example, it was shown in \citet[Corollary ~2]{zhang2013communication} that, under appropriate regularity conditions,
\begin{equation}\label{eq:OS_bound}
	\E \big\|\bar{\bds{\theta}} -\bds{\theta}^*\big\|_2^2 \le \frac{C_1}{ N} + \frac{C_2}{n^2} +O\Bigg( \frac{1}{ Nn} + \frac{1}{n^3} \Bigg),
\end{equation}
where $ C_1,\ C_2 $ are some positive constants. If $ n $ is sufficiently large such that $n^{-2}= o(N^{-1})$, then the dominant term in \eqref{eq:OS_bound} becomes $ C_1/N $, and is of the order $ O(N^{-1})$. It is the same as that of the whole sample estimator. This also implies that, in order to obtain the global convergence rate, we should not divide the whole sample into too many parts. A further improved theoretical results were obtained by \cite{rosenblatt2016optimality}. They showed that the one-shot estimator is first order equivalent to the whole sample estimator. However, the second-order error terms of $\bar{\bds{\theta}}$ can be non-negligible for nonlinear models. Similar observation was also obtained by \cite{huang2015distributed}. The work of \cite{duchi2014optimality} revealed that the minimal communication budget to attain the global estimation error for linear regression is $\mathcal{O}(Kp)$ bits up to a logarithmic factor under some conditions. This result matches the simple averaging procedure and confirms the sharpness of the bound in \eqref{eq:OS_bound}.  To further reduce the bias, a novel subsampling method was developed by \cite{zhang2013communication}. By this technique, the error bound is improved to be $O(N^{-1}+n^{-3})$, which relaxes the restriction on the number of machines.

Instead of the linear combination of local maximum likelihood estimates (MLEs) as simple average, \cite{liu2014distributed} proposed a KL-divergence based combination method,
\begin{align*}
	\hat{\bds\theta}_{\text{KL}} = \argmin_{\bds\theta\in\Theta}\sum_{k=1}^K \operatorname{KL}\Big(p(x|\hat{\bds\theta}_k)\  \big\|\ p(x|\bds\theta)\Big),
	\label{eq:KL_based}
\end{align*}
where $p(x|\bds\theta)$ is the probability density of $\mathbb{P}_{\bds\theta}$ with respect to some proper measure $\mu$ and KL-divergence is defined by $\operatorname{KL}\big(p(x)\  \big\|\ q(x)\big)=\int_{\mathcal{X}} p(x)\ \log\{{p(x)}/{q(x)}\} d \mu (x)$.
It was shown that $\hat{\bds\theta}_{\text{KL}}$ is exactly the global MLE $\hat{\bds\theta}$ if $\{\mathbb{P}_{\bds{\theta}}: \bds{\theta} \in \Theta \}$ is a full exponential family (defined in their paper). This sheds light on the inference about generalized linear models (GLMs) based on exponential likelihood.

In many cases, some local machines might suffer from data of poor quality. This could lead to abnormal local estimates, which further degrade the statistical efficiency of the final estimator. To fix the problem, \cite{minsker2019distributed} devised a robust assembling method. It leads to an estimator as
$ \hat{\bds{\theta}}_\text{robust} = \argmin_{\bds{\theta}\in\Theta}\sum_{k=1}^K \rho(|\bds{\theta} -  \hat{\bds{\theta}}_k|)$,
where $\rho(\cdot)$ is a robust loss function satisfying some conditions. For example, when $\rho(u)=u$ and $ p=1 $ (univariate case), $ \hat{\bds{\theta}}_\text{robust} $ is the median of $\hat{\bds{\theta}}_k$'s. It should be more robust against outliers as compared with the simple average. Under some regularity conditions, they showed that $ \hat{\bds{\theta}}_\text{robust} $ achieves the same convergence rate as the whole sample estimator provided $K\le O(\sqrt{N})$.

\subsection{Iterative approach}

Although one-shot approach involves little communication cost, it suffers from several disadvantages. First, the local machines need to have sufficient amount of data (e.g., $n\gg \sqrt{N}$). Otherwise the aggregated estimator cannot reach the convergence rate as the global estimator. This prevents us from utilizing many machines to speed up the computation process \citep{wang2017efficient, jordan2019communication}. Second, the simple averaging estimator is often poor in performance for nonlinear models \citep{rosenblatt2016optimality,huang2015distributed,jordan2019communication}. Last, when $p$ is diverging with $N$, the situation could be even worse \citep{rosenblatt2016optimality, lee2017communication}. This suggests that carefully designed algorithms allowing a reasonable number of iterations should be useful for a distributed system.

Inspired by the one-step method in the $M$-estimator theory, \cite{huang2015distributed} proposed an one-step refinement of the simple averaging estimator. Let us recall that $ \bar{\bds{\theta}}$ is the one-shot averaging estimator. To further improve its statistical efficiency, it should be broadcast to each local machine. Next, local gradient $\nabla \mathcal{L}_k(\bar{\bds{\theta}}) $ and local Hessian $ \nabla^2 \mathcal{L}_k(\bar{\bds{\theta}}) $ can be computed on each $ \mathcal{M}_k $. Then, they are reported to $ \mathcal{M}_\text{center} $ to form the central gradient $ \nabla \mathcal{L}(\bar{\bds{\theta}}) = K^{-1} \sum_{k=1}^K \nabla \mathcal{L}_k(\bar{\bds{\theta}})  $ and Hessian $ \nabla^2 \mathcal{L}(\bar{\bds{\theta}}) = K^{-1} \sum_{k=1}^K \nabla^2 \mathcal{L}_k(\bar{\bds{\theta}})$. Thus an one-step updated estimator can be constructed on $ \mathcal{M}_\text{center} $ as
\begin{equation}\label{eq:one_step}
	\hat{\bds{\theta}}^{(1)}= \bar{\bds{\theta}} - [\nabla^2 \mathcal{L}(\bar{\bds{\theta}})]^{-1} \nabla \mathcal{L}(\bar{\bds{\theta}}).
\end{equation}
Compared with one-shot estimator, $  \hat{\bds{\theta}}^{(1)} $ involves one more round of communication cost. Nevertheless, the statistical efficiency of the resulting estimator could be well improved. In fact, \cite{huang2015distributed} showed that
\begin{equation*}
	\E \big\| \hat{\bds{\theta}}^{(1)}  -\bds{\theta}^*\big\|_2^2\le \frac{C_1}{N} + O\Bigg(\frac{1}{n^4} + \frac{1}{N^2} \Bigg),
\end{equation*}
where $ C_1 >0$ is some constant. Obviously, this is a lower upper bound of mean squared error than that in \eqref{eq:OS_bound}. To attain the global convergence rate, the local sample size should satisfy $ n^{-4} = o(N^{-1})$, which is a much milder condition. Furthermore, they showed that $ \hat{\bds{\theta}}^{(1)} $ also has the same asymptotic efficiency as the whole sample estimator $ \hat{\bds\theta} $ under some regularity conditions. 

A natural idea to further extend the one-step estimator is to allow the iteration \eqref{eq:one_step} to be executed many times. Specifically, let $ \hat{\bds{\theta}}^{(t)} $ be the estimator of the $ t $-th iteration. Then, we can use \eqref{eq:one_step} by replacing $ \bar{\bds\theta} $ with $ \hat{\bds{\theta}}^{(t)} $ to generate the next step estimator $ \hat{\bds{\theta}}^{(t+1)} $ (see Figure \ref{fig:oneshot and iteative}(b)).
However, this requires a large number of Hessian matrices to be computed and transferred. If the parameter dimension $ p $ is relatively high, this will lead to a significant communication cost of the order $ O(Kp^2) $. To fix the problem, \cite{shamir2014communication} proposed an approximate Newton method, which conducts Newton-type iteration distributedly without transferring the Hessian matrices. 
Following this strategy, \cite{jordan2019communication} developed an approximate likelihood approach. Their key idea is to update Hessian matrix on one single machine (e.g., $ \mathcal{M}_\text{center} $) only. Then, \eqref{eq:one_step} can be revised to be
\begin{equation*}
	\hat{\bds{\theta}}^{(t+1)}=\hat{\bds{\theta}}^{(t)}- \Big[\nabla^2 \mathcal{L}_\text{center}(\hat{\bds{\theta}}^{(t)} )\Big]^{-1} \nabla \mathcal{L}(\hat{\bds{\theta}}^{(t)} ),
\end{equation*}
where $ \nabla^2 \mathcal{L}_\text{center} $ is the Hessian matrix computed on the central machine.
By doing so, the communication cost due to transmission of Hessian matrices can be saved. Under some conditions, they showed that
\begin{equation}\label{eq:linear_convergence}
	\| \hat{\bds{\theta}}^{(t+1)}  - \hat{\bds\theta}\|_2 \le \frac{C_1}{\sqrt{n}} \| \hat{\bds{\theta}}^{(t)}  - \hat{\bds\theta}\|_2, \quad \text{for } t\ge 0,  
\end{equation}
holds with high probability, where $ C_1 >0$ is some constant. By the linear convergence of the estimates \eqref{eq:linear_convergence}, we can see that it requires $ [\log K/ \log n] $ iterations to achieve the $ \sqrt{N} $-consistency as the whole sample estimator $ \hat{\bds\theta} $, provided $ \hat{\bds{\theta}}^{(0)}  $ is $ \sqrt{n} $-consistent. Note that if $ n =K= \sqrt{N} $, one iteration suffices to attain the optimal convergence rate. However, the satisfactory performance of this method relies on a good choice of the machine, on which the Hessian needs to be updated \citep{fan2019communication}. To fix the problems, \cite{fan2019communication} added an extra regularized term to the approximate likelihood used in \cite{jordan2019communication}. With this modification, the performance of the resulting estimator can be well improved. Theoretically, they showed a similar linear convergence rate under some more general conditions, which require no strict homogeneity of the local loss functions.

\subsection{Shrinkage methods}
We study various shrinkage methods for sparse estimation in this subsection. 
For a high-dimensional problem, especially when the dimension of $\bds\theta^*$ is larger than the sample size $ N $, it is difficult to estimate $\bds\theta^*$ without any additional assumptions \citep{hastie2015statistical}. A popular constraint for tackling these problems is sparsity, which assumes only a subset of the entries in $\theta^*$ is non-zero. The index of non-zero entries is called the support of $\bds\theta^*$, that is
\begin{equation*}
	\operatorname{supp}(\bds\theta^*)=\mathcal{A}^* = \Big\{1\le j\le p:\ \theta^*_j \ne 0\Big\}.
\end{equation*}
To induce a sparse solution, an additional regularization term of $\bds\theta$ is often introduced in the loss function. Specifically, we need to solve the shrinkage regression problem as $ \min_{\bds{\theta}\in\Theta}\{ \mathcal{L}(\bds{\theta})+ \sum_{j=1}^{p}\rho_\lambda(|\theta_j|) \} $, where $\rho_\lambda(\cdot)$ is a penalty function with a regularization parameter $\lambda> 0$. Popular choices are LASSO \citep{tibshirani1996regression}, SCAD \citep{fan2001variable} and others discussed in \cite{zhang2012general}. For simplicity, we consider the LASSO estimator in the framework of the linear model. Specifically, the whole sample estimator is computed as
\begin{equation*}
	\hat{\bds{\theta}}_\lambda=\argmin_{ \bds{\theta}\in\Theta} \Big\{ \frac{1}{N} \| \bds Y - \bds X \bds\theta\|_2^2  + \lambda \|\bds\theta\|_1 \Big\},
\end{equation*}
where $\bds Y=(Y_1,\cdots,Y_N)^\top \in\R^N$ is the vector of response, $\bds X=(X_1,\dots,X_N)^\top \in\R^{N\times p}$ is the design matrix, and $ \|\bds\theta\|_1 = \sum_{j=1}^p|\theta_j| $ denotes the $ l_1 $-norm of $ \bds\theta $. It is known that the LASSO procedure would produce biased estimators for the large coefficients. This is undesirable for the simple average procedures, since average cannot eliminate the systematic bias. To reduce bias, \cite{javanmard2014confidence} proposed a debiasing technique for the lasso estimator, that is
\begin{equation}
	\hat{\bds \theta}^\text{(d)}_\lambda= \hat{\bds\theta}_\lambda + \frac{1}{N} \bds M \bds X^\top \Big(\bds Y - \bds X \hat{\bds\theta}_\lambda\Big),
	\label{eq:debiased}
\end{equation}
where $\bds M\in\R^{p\times p}$ is an approximation to the inverse of $\hat{\Sigma} = \bds{X}^\top \boldsymbol{X}/N$. It appears that when $\hat{\Sigma}$ is invertible (e.g., when $ N\gg p $), setting $\bds M=\hat{\Sigma}^{-1}$ gives $ \hat{\bds\theta}^\text{(d)}_\lambda =(\bds X^\top \bds X)^{-1}\bds X^\top \bds Y $, which is the ordinary least squares estimator and obviously unbiased. Hence, procedure \eqref{eq:debiased} compensates for the bias incurred by $\ell_1$ regularization in some sense. 

By this debiasing technique, \cite{lee2017communication} developed an one-shot type estimator for the LASSO problem. Specifically, let $\hat{\bds{\theta}}_{k,\lambda}^\text{(d)} $ be the debiased LASSO estimator computed on $ \mathcal{M}_k $. Then an averaging estimator can be constructed on $  \mathcal{M}_\text{center}$ as $ \bar{\bds{\theta}}_\lambda=K^{-1}\sum_{k=1}^K  \hat{\bds{\theta}}_{k,\lambda}^\text{(d)}$. Unfortunately, the sparsity level can be seriously degraded by averaging. For this reason, a hard threshold step often comes as a remedy. It was noticed that the debiasing step is computationally expensive. Hence an improved algorithm was also proposed to alleviate the computational cost of this step. Under certain conditions, they showed that the resulting estimator has the same convergence rate as the whole sample estimator. \cite{battey2018distributed} investigated the same problem with additional study on hypothesis testing. Furthermore, a refitted estimation procedure was used to preserve the global oracle property of the distributed estimator. An extension to high dimensional GLMs can also be found in \cite{lee2017communication} and \cite{battey2018distributed}. For this model, \cite{chen2014split} implemented a majority voting method to combine the regularized local estimates without a debiasing step. For the low dimensional sparse problem with smooth loss function (e.g., GLMs, Cox model), \cite{zhu2019least} developed a local quadratic approximation method with an adaptive-LASSO type penalty. They showed rigorously that the resulting estimator can be as good as the global oracle estimator. 

Intuitively, above one-shot methods may need a stringent condition on the local sample size to meet the global convergence rate due to the limited communication. In fact, the simple averaging estimator requires $n\ge O(Ks^2\log p)$ to match the oracle rate in the context of sparse linear model \citep{lee2017communication}, where $s=|\mathcal{A}^*|$ is the number of non-zero entries of $\theta^*$. For this problem, \cite{wang2017efficient} and \cite{jordan2019communication} independently proposed a communication-efficient iterative algorithm, which constructs a regularized likelihood by using local Hessian matrix. As demonstrated by \cite{wang2017efficient}, the one-step estimator $\hat{\bds\theta}^{(1)}$ suffices to achieve the global convergence rate if $n\ge O(Ks^2\log p)$ \citep[the condition used in ][]{lee2017communication}. Furthermore, if multi-round communication is allowed, $\hat{\theta}^{(t+1)}$ (i.e., estimator of the $ (t+1) $-th iteration) can match the estimator based on the whole sample as long as $n\ge O(s^2\log p)$ and $t\ge O(\log K)$, under some certain conditions. 

\subsection{Non-smooth loss based models}

The methods we described above typically require the loss function $\mathcal{L}$ to be sufficiently smooth, although a non-smooth regularization term is permitted \citep[see e.g.,][]{zhang2013communication, wang2017efficient, jordan2019communication, zhu2019least}. However, there are also some useful methods involving non-smooth loss functions, such as quantile regression and support vector machine. It is then of great interest to develop distributed methods for these methods. 

We first focus on the quantile regression (QR) model. The QR model has a widespread use in statistics and econometrics, and performs more robust against the outliers than the ordinary quadratic loss \citep{koenker2005quantile}. Specifically, a QR model assumes $ Y_i = X_i^\top \bds{\theta}^* + \varepsilon_i,\, i \in \mathbb{S}$,
where $X_i\in\R^{p}$ is the covariate vector, $ Y_i $ is the corresponding response, $\bds{\theta}_\tau^*\in\R^p$ is the true regression coefficient, and $\varepsilon_i$ is the random noise satisfying $\mathbb{P}(\epsilon_i\le 0 | X_i)=\tau $, where $\tau\in (0,1)$ is a known quantile level. It is known that $\bds{\theta}_\tau^*$ is the minimizer of $\E[\rho_\tau(Y_i - X_i^\top  \bds{\theta})] $. Here $\rho_\tau(u)=u(\tau - \mathbf{1}\{u\le 0\}) = u(\mathbf{1}\{u> 0\} +\tau-1)$ is the non-differentiable check-loss function, where $ \mathbf{1}\{\cdot\} $ is the indicator function. When data size $N$ is moderate, we can estimate $\bds{\theta}_\tau^*$ by $\hat{\bds\theta}_\tau = \min_{\bds{\theta}\in\Theta} N^{-1} \sum_{i \in \mathbb{S}}\rho_\tau(Y_i-X_i^\top \bds{\theta} )$ on one single machine. However, when $N$ is very large, a distributed system has to be used. Accordingly, distributed estimators have to be developed. 

\begin{figure}[h]
	\centering
	\subfloat[Approximation of QR loss with $\tau = 0.6$.]{%
		\resizebox*{0.45\textwidth}{!}{\includegraphics{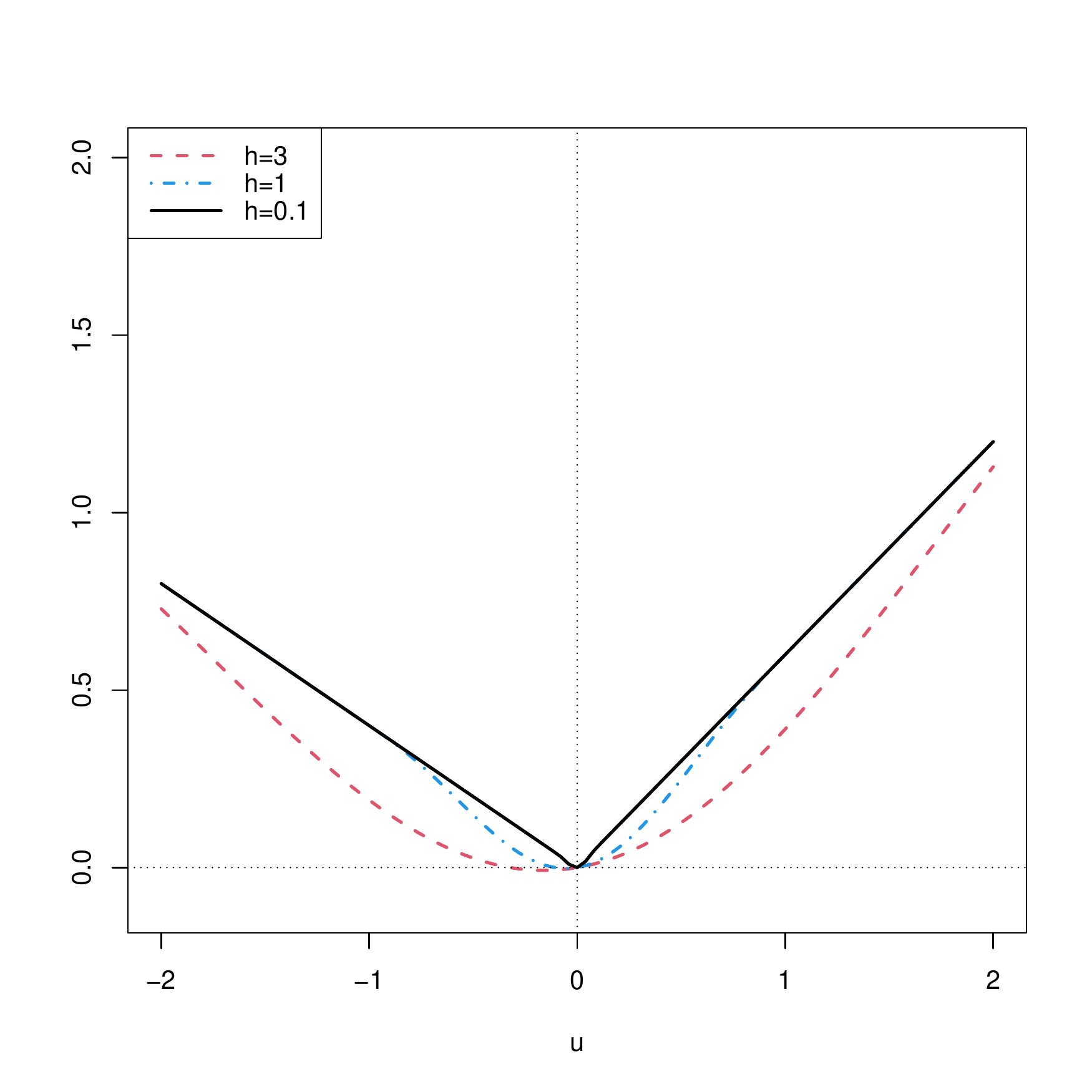}}}\hspace{5pt}
	\subfloat[Approximation of hinge loss.]{%
		\resizebox*{0.45\textwidth}{!}{\includegraphics{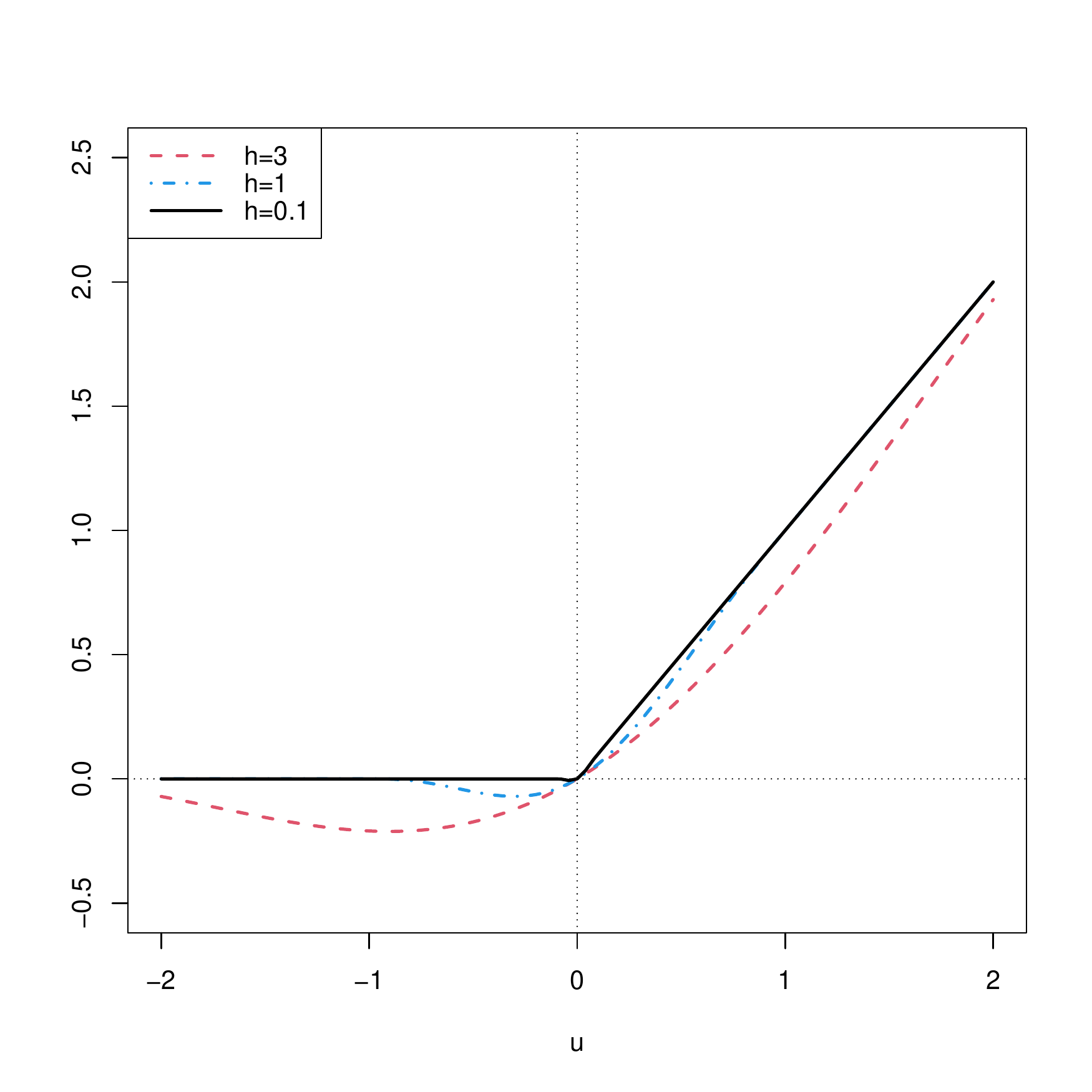}}}
	\caption{Approximation of two non-smooth loss functions.} \label{fig:loss_approx}
\end{figure}

In this regard, \cite{volgushev2019distributed} studied the one-shot averaging type estimator. Specifically, a local estimator $ \hat{\bds\theta}_{k, \tau} $ is first computed on each local machine $ \mathcal{M}_k $. Then, the averaging estimator is assembled as $  \bar{\bds\theta}_{\tau} =K^{-1}\sum_{k=1}^K  \hat{\bds\theta}_{k, \tau}  $ on the central machine $ \mathcal{M}_\text{center} $. They further investigated the theoretical properties of the averaging estimator in detail. It was shown that the if the number of machines satisfies $ K = o(\sqrt{N}/\log N) $, then $ \bar{\bds\theta}_{\tau}  $ should be as efficient as the whole sample estimator $ \hat{\bds\theta}_{\tau}  $ under some regularity conditions. \cite{chenlanjue2020quantile} proposed an estimating equation based one-shot approach for the QR problem. The asymptotic equivalence between the resulting estimator and the whole sample estimator was also established under $ K=o(N^{1/4}) $ and some other conditions. It can be seen that the performance of one-shot approaches relies more on the local sample size. In fact, \cite{volgushev2019distributed} showed that $ K = o(\sqrt{N}) $ is a necessary condition for the global efficiency of the simple averaging estimator $ \bar{\bds\theta}_{\tau} $. To remove the constraint $ K = o(\sqrt{N}) $ on the number of machines, \cite{chen2019quantile} proposed a iterative approach. Their key idea is to approximate the check-loss function by a smooth alternative. More specifically, they approximated $\mathbf{1}\{u> 0\}$ by a smooth function  $H(u/h)$, where $H(\cdot)$ is a smooth cumulative distribution function and $ h > 0 $ is the tuning parameter controlling the approximation goodness (see Figure \ref{fig:loss_approx}(a)). With this modification, the algorithm can update the estimates by
\begin{equation}
	\hat{\bds\theta}_\tau^{(t+1)} =\big[\bds V(\hat{\bds\theta}_\tau^{(t)})\big]^{-1} \bds U(\hat{\bds\theta}_\tau^{(t)} ),
	\label{eq:qr_iter}
\end{equation}
where $\bds U(\bds\theta) = \sum_{k=1}^K \bds U_k(\bds\theta)$, $\bds V(\bds\theta)= \sum_{k=1}^K \bds V_k(\bds\theta)$, and $ \bds U_k\in \R^{p}, \ \bds V_k\in \R^{p\times p} $ depend only on the bandwidth $h$ and local sample $ \mathcal{S}_k $. It was shown that a constant number of rounds of iteration suffices to match the convergence rate of the  whole sample estimator. Thus, the communication cost is roughly of the order $O(K p^2\big)$, which is not applicable when $p$ is very large. For the high dimensional QR problem, \cite{zhao2014general} and \cite{zhao2019debiasing} adopted an one-shot averaging method based on the debiased local estimates as that in \eqref{eq:debiased}. Accordingly, \cite{chen2020distributed} proposed a communication-efficient multi-round algorithm inspired by the approximate Newton method \citep{shamir2014communication}. This iterative approach removes the restriction on the number of machines. A revised divide-and-conquer stochastic gradient descent method for QR and other models with diverging dimension can be found in \cite{chen2021first}.

We next consider the support vector machine (SVM), which is one of the most successful statistical learning methods \citep{vapnik2013nature}. The classical SVM is aimed at the binary classification problem, i.e., the response variable $Y_i\in\{-1,1\}$. Formally, a standard linear SVM solves the problem $\hat{\bds \theta}_\lambda = \argmin_{\bds \theta \in \Theta}  N^{-1}\sum_{i\in \mathbb{S}} \big(1- Y_i X_i^\top \bds \theta\big)_+ +\lambda\|\bds \theta\|_2^2$,
where  $(u)_+=u\mathbf{1}(u>0)$ is the hinge loss, and $\lambda>0$ is the regularization parameter.
By the same smooth technique used in \cite{chen2019quantile}, i.e., replacing the hinge loss with a smooth alternative (see Figure \ref{fig:loss_approx}(b) ), \cite{wang2019distributed} proposed an iterative algorithm like \eqref{eq:qr_iter}. To reduce the communication cost incurred by transferring matrices, they further employed the approximate Newton method \citep{shamir2014communication}.  Theoretically, they showed the asymptotic normality of the estimator, which can be used to construct confidence interval. For the ultra-high dimensional SVM problem, \cite{lian2018divide} studied the one-shot averaging method with debiasing procedure similar to \eqref{eq:debiased}.

\section{Nonparametric models}

Different from parametric models, a nonparametric model typically involves infinite-dimensional parameters. In this section, we focus mainly on the nonparametric regression problems. Specifically, consider here a general regression model as $Y_i = f^*(X_i)+\varepsilon_i,\, i\in\mathbb{S}$, where $f^*(\cdot)$ is an unknown but sufficiently smooth function and $\varepsilon_i$ is the random noise with zero mean. The aim of nonparametric regression is to estimate function $f^*\in\mathcal{F}$, where $\mathcal{F}$ is a given nonparametric class of functions.

\subsection{Local smoothing}
One way to estimate $f^*(\cdot)$ is to fit a locally constant model by kernel smoothing \citep{fan1996local}. More concretely, the whole sample estimator is given by
\begin{equation*}
	\hat{f}_h(x) = \sum_{i \in \mathbb{S}} W_{h,X_i}(x)Y_i,
\end{equation*}
where the $ W_{h, X_i}(x)\ge 0 $ is the local weight at $ X=x $ satisfying $\sum_{i\in \mathbb{S}} W_{h, X_i}(x)=1$. Specifically, for a Nadaraya-Watson kernel estimator, we should have $ W_{h, X_i}(x) = K\big((X_i-x)/h\big) /  \sum_{i' \in \mathbb{S}} K\big((X_{i'}-x)/h\big)$, where $ K(\cdot) $ is a kernel function and $ h>0$ is the bandwidth. In the univariate case ($ p=1 $), classical theory stated that the mean squared error of $ \hat{f}_h (x) $ is of the order $O(h^4 + (Nh)^{-1})  $ \citep{fan1996local}. Thus, the optimal rate $ O(N^{-4/5}) $ can be achieved by choosing bandwidth $ h =O (N^{-1/5}) $.

For a distributed kernel smoothing, an one-shot estimator can also be constructed. Let $  \hat{f}_{k,h}(x) $ be the local estimator computed on $ \mathcal{M}_k $. Then an averaging estimator can be obtained as $ \bar{f}_h(x)=K^{-1}\sum_{k=1}^K \hat{f}_{k,h}(x)$.
\cite{chang2017divide} studied the theoretical properties of $ \bar{f}_h(x) $ in a specific function space $ \mathcal{F} $. They established the same minimax convergence rate of $ \bar{f}_h(x) $ as that of the whole sample estimator. However, they found that a strict restriction on the number of machines $ K $ is needed to achieve this optimal rate. To fix the problem, two solutions were provided. They are, respectively, a date-dependent bandwidth selection algorithm and an algorithm with a qualiﬁcation step.

Nearest neighbors method can be regarded as another local smoothing method. \cite{qiao2019rates} studied the Nearest neighbors classification in a distributed setting, where the optimal number of neighbors to achieve the optimal rate of convergence was derived. \cite{li2013statistical} discussed the problem of density estimation for scattered datasets. \cite{kaplan2019optimal} focused on the choice of bandwidth for nonparametric smoothing techniques. All the works above in this subsection indicate that the bandwidth (or local smoothing parameter) used in the distributed setting should be adjusted according to the whole sample size $N$, other than the local sample size $n$.

\subsection{RKHS methods}
\label{subsec:kernel}

We next discuss another popular nonparametric regression method. This is reproducing kernel Hilbert space (RKHS) method. An RKHS $ \mathcal{H} $ can be induced by a continuous, symmetric and positive semi-definite kernel function $ \mathcal{K}(\cdot, \cdot): \R^{p}\times \R^{p}\mapsto \R $. Two typical examples are: polynomial kernel $\mathcal{K}(x_1,x_2)=(x_1^\top x_2+1)^d$ with an integer $d\ge1$, and radical kernel $\mathcal{K}(x_1,x_2)=\exp(-\gamma\|x_1-x_2\|_2^2)$ with $\gamma > 0$. Refer to, for example, \cite{wahba1990spline, berlinet2011reproducing} for more details about RKHS. Then, our target is to find an $ \hat{f} \in \mathcal{H}$ so that the following penalized empirical loss can be minimized. That leads to the whole sample estimator as
\begin{equation}\label{eq:krr}
	\hat{f}_\lambda  =\argmin_{f\in \mathcal{H}}\Big\{\frac{1}{N}\sum_{i \in \mathbb{S}} (Y_i-f(X_i))^2 +\lambda \|f\|_{\mathcal{H}}^2  \Big\},
\end{equation}
where $  \|\cdot\|_{\mathcal{H}} $ is the norm associated with the RKHS $ \mathcal{H} $ and $ \lambda>0 $ is the regularization parameter. This problem is also known as kernel ridge regression (KRR). By the representer theorem for the RKHS \citep{wahba1990spline}, any solution to the problem \eqref{eq:krr} must have the linear form as $ \hat{f}_\lambda(x) = \sum_{i\in \mathbb{S}} \alpha_i \mathcal{K}(X_i, x) $, where $ \alpha_i \in \R $ for each $ i \in \mathbb{S} $. By this property, we can treat the KRR as a parametric problem with unknown parameter $ \alpha = (\alpha_1,\dots, \alpha_N)^\top\in \R^{N} $. The error bounds of the whole sample estimator $ \hat{f}_\lambda $ has been well established in the existing literature \citep[e.g.,][]{zhang2005learning, steinwart2009optimal}. However, a standard implementation of the KRR involves inverting a kernel matrix in $\R^{N\times N}$ \citep{saunders1998ridge}. Therefore, when $ N $ is extremely large, it is time consuming or even computationally infeasible to process the whole sample on a single machine. Thus, we should consider a distributed system.

In this regard, \cite{zhang2015divide} studied the distributed KRR by taking the one-shot averaging approach. Specifically, each machine $\mathcal{M}_k$ computes local KRR estimate $\hat{f}_{k,\lambda} $ by \eqref{eq:krr} based on local sample $\mathcal{S}_k$. Then the central machine $\mathcal{M}_{\text{center}}$ averages them to obtain final estimator $\bar{f}_\lambda = K^{-1} \sum_{k=1}^{K}\hat{f}_{k,\lambda} $. Theoretically, they established the optimal convergence rate of mean squared error for $ \bar{f}_\lambda $ with different types of kernel functions, under some regularity conditions. \cite{lin2017distributed} derived a similar optimal error bound under some more relaxed conditions. \cite{xu2016feasibility} extended the loss function in \eqref{eq:krr} to a further general form. Some related works on the distributed KRR problem by one-shot averaging approach can be found in \cite{shang2017computational, lin2018distributed, mucke2018parallelizing, guo2019distributed, wang2019sharper} and many others. It was noted that these one-shot approaches require the number of machines diverges in a relative slow speed to meet the global convergence rate. To fix the problem, \cite{chang2017distributed} proposed a semi-supervised learning framework by utilizing the additional unlabeled data (i.e., observations without response $ Y_i $). Latest work of \cite{lin2020distributed} allowed communication between machines for the distributed KRR problem to improve the performance. In order to choose an optimal tuning parameter $ \lambda $ in \eqref{eq:krr}, \cite{xu2018optimal} proposed a distributed generalized cross-validation method. 

For semiparametric models, \cite{zhao2016partially} considered a partially linear model with heterogeneous data in a distributed setting. Specifically, they assumed the following model
\begin{equation} \label{eq:partially_linear}
	Y_i = X_i^\top \bds\theta^*_{(k)} + f^*(W_i)+ \varepsilon_i, \quad i\in \mathcal{S}_k,
\end{equation}
where $ W_i\in \R $ is an additional covariate, $ f^*(\cdot) $ is the unknown function, and $ \bds\theta^*_{(k)} \in \R^{p}$ is the true linear coefficient associated with the data on $ \mathcal{M}_k $ for $ 1 \le k\le K $. In other words, the local data on different machines are assumed to share the same nonparametric part, but are allowed to have different linear coefficients. To estimate the unknown function and coefficients, they extended the classical RKHS theory to cope with the partially linear function space. Under some regularity conditions, the resulting estimator of the nonparametric part is shown to be as efficient as the whole sample estimator, provided the number of machines does not grow too fast. The case of high dimensional linear part was also investigated. For example, under the homogeneity assumption (i.e., the linear coefficients $\bds \theta^*_{(k)} $'s are assumed to be identical to $ \bds \theta^* $ across different machines), \cite{lv2017debiased} adopted the one-shot averaging approach with debiasing technique analogous to \eqref{eq:debiased} to estimate the linear coefficient. \cite{lian2019projected} considered the same heterogeneous model as in \eqref{eq:partially_linear}, but the linear part is assumed in a high dimensional setting (i.e., $ p>N $). For this model, they proposed a novel projection approach to estimate the common nonparametric part (not in an RKHS framework). Theoretically, the asymptotic normality of the one-shot averaging estimator for the nonparametric function was established under some certain conditions.

\section{Other related works}

\subsection{Principal component analysis}

Principal component analysis (PCA) is a common procedure to reduce the dimension of the data. It is widely used in the practical data analysis. Unlike the regression problems, PCA is an unsupervised method, which does not require a response variable $Y$. To conduct a PCA, a covariance matrix $\hat{\Sigma}$ needs to be constructed as $ \hat{\Sigma}=N^{-1}\sum_{i \in \mathbb{S}}X_i X_i^\top  $, where $ X_i $'s are assumed to be centralized already. Next, a standard singular value decomposition (SVD) is applied to $ \hat{\Sigma} $. That leads to $ \hat{\Sigma}=\hat{V}\hat{D}\hat{V}^\top$, where $ \hat{D} $ is a diagonal matrix of eigenvalues and $ \hat{V} $ is an orthogonal matrix of eigenvectors. Then, the columns of $ \hat{V} $ are the principal component directions that we need.

In a distributed setting, simple average of the eigenvectors estimated locally cannot give a valid result. To solve the problem, \cite{fan2019distributed} developed a divide-and-conquer algorithm for estimating eigenspaces. It involves only one single round of communication. This algorithm is quite easy to implement as well. We state it as follows \citep[Algorithm~1]{fan2019distributed}:

\begin{enumerate}
	\item For each $k=1,\cdots,K$, machine $\mathcal{M}_k$ computes $d$ leading eigenvectors of the local sample covariance matrix $\hat{\Sigma}_{k}=n^{-1} \sum_{i\in \mathcal{S}_k} X_i X_i^\top $, denoted by $ \hat{v}_{1,k},\cdots,\hat{v}_{d,k} \in\R^p$. Next, they are arranged by columns in $\hat{V}_k=(\hat{v}_{1,k},\cdots,\hat{v}_{d,k})\in \R^{p\times d}$, which is then sent to the central machine $\mathcal{M}_{\text{center}}$.
	
	\item The central machine $\mathcal{M}_{\text{center}}$ averages $K$ local projection matrices to obtain
	$\tilde{\Sigma}=K^{-1} \sum_{k=1}^K \hat{V}_{k} \hat{V}_{k}^\top$.
	Then it computes $d$ leading eigenvectors of $ \tilde{\Sigma} $, denoted by
	$\tilde{v}_1, \cdots,\tilde{v}_{d} \in \R^p$. The $ \tilde{v}_1, \cdots,\tilde{v}_{d}$ are the estimators of the first $ d $ principal component directions that we need.
\end{enumerate}
It is noticeable that the communication cost of above one-shot algorithm is of the order $O(Kdp)$. This can be considered to be communication-efficient since $d$ is usually much smaller than $p$ in practice.
\cite{fan2019distributed} showed that, under some appropriate conditions, the distributed estimator achieves the same convergence rate as the global estimator. The cases of heterogeneous local data were also investigated in their work. To further remove the restriction on the number of machines, \cite{chen2021distributed} proposed a communication-efficient multi-round algorithm based on the approximate Newton method \citep{shamir2014communication}.

\subsection{Feature screening}
Massive datasets often involve ultrahigh dimensional data, for which feature screening is critically important \citep{fan2008sure}. To fix the idea, consider a standard linear regression model as $ Y_i = X_i^\top \bds{\theta}^* + \epsilon_i,\ i \in \mathbb{S} $, where $X_i\in\R^{p}$ is the covariate vector, $ Y_i $ is the corresponding response, $\bds{\theta}^*\in\R^p$ is the true parameter, and $\varepsilon_i$ is the random noise. To screen for the most promising features, the seminal method of sure independence screening (SIS) has been proposed by \cite{fan2008sure}. Specifically, let $ \mathcal{A}^* = \{1\le j \le p: \theta_j^* \ne 0 \} $ be the true sparse model. Let $ \omega_j $ be the Pearson correlation between $ j $th feature and response $ Y $. Then, SIS screens features by a hard threshold procedure as $\hat{\mathcal{A}}_\gamma=\{1\le j \le p: |\hat{\omega}_j| > \gamma \}$,
where $ \gamma $ is a prespecified threshold and $ \hat{\omega}_j $ is the whole sample estimator of $ \omega_j $. Under some specific conditions, \cite{fan2008sure} showed the sure screening property for SIS, that is,
\begin{align*}
	\mathbb{P} ({\mathcal{A}^*\subset \hat{\mathcal{A}}_\gamma}) \to 1\quad \text{as } N\to\infty.
\end{align*}

However, the estimator $ \hat{\omega}_j $ is usually biased for many correlation measures. This indicates that a direct one-shot averaging approach is unlikely to be the best practice for the distributed system. To fix the problem, \cite{li2020distributed} proposed a novel debiasing technique. They found that many correlation measures can be expressed as $ \omega_j = g(\nu_{1},\dots,\nu_{s}) $, including Pearson correlation used above, Kendall $\tau$ rank correlation, SIRS correlation \citep{zhu2011model}, etc. Therefore, they used $ U $-statistics to estimate the components $ \nu_{q}, 1\le q\le s $ on local machines. Then, these unbiased estimators of $ \nu_{q} $'s given by local machines are averaged on the central machine $ \mathcal{M}_\text{center} $. Consequently, $ \mathcal{M}_\text{center} $ can construct distributed estimator $ \tilde{\omega}_j  $ by the averaging estimators of the components in the known function $ g $. Finally, they showed the sure screening property of $ \hat{\mathcal{A}}_\gamma $ based on the distributed estimators under some regularity conditions. When the feature dimension is much larger than the sample size (i.e., $ p\gg N $), another distributed computing strategy is to partition the whole data by features, other than by samples. Refer to, for example, \cite{song2015split, yang2016feature} for more details.

\subsection{Bootstrap}

Bootstrap and related resampling techniques provide a general and easily implemented procedure for automatically statistical inference. However, these methods are usually computationally expensive. When sample size $N$ is very large, it would be even practically infeasible to conduct. To mitigate this computing issue, various alternative methods have been proposed, such as subsamping approach \citep{politis1999subsampling} and ``$m$-out-of-$n$'' bootstrap \citep{bickel2012resampling}. Their key idea is to reduce the resample size. However, due to the difference between the size of whole sample and resample, an additional correction step is generally required to rescale the result. This makes these methods less automatic.

To solve this problem, \cite{kleiner2014scalable} proposed the bag of little bootstraps (BLB) method. It integrates the idea of subsampling and can be computed distributedly without a correction step. Suppose that $N$ sample units have been randomly and evenly partitioned to $K$ machines. Consider that we want to assess the accuracy of the point estimator for some parameter $ \bds{\theta} $. Then we summarize their algorithm as follows.
\begin{enumerate}
	\item For each $1\le k\le K$, machine $\mathcal{M}_k$ draws $r$ samples of size $N$ (instead of $n$) from $\mathcal{S}_k$ with replacement. Then it computes $ r $ estimates of $\bds{\theta} $ separately based on the $ r $ resamples drawn above. After that, each $\mathcal{M}_k$ computes some accuracy measure, denoted by $\hat{\xi}_k$ (e.g., confidence region), by the $r$ estimates above. Finally, all of the local machines send $\hat{\xi}_k$'s to the central machine $\mathcal{M}_{\text{center}}$.
	
	\item The central machine $\mathcal{M}_{\text{center}}$ aggregates these $ \hat{\xi}_k $'s by $\bar{\xi}=K^{-1}\sum_{k=1}^K \hat{\xi}_k$. The $ \bar{\xi} $ is the final accuracy measure that we need.
\end{enumerate}
It is remarkable that one does not need to process datasets of size $ N $ on local machines actually, although the nominal size of resample is $N$. This is because each machine contains at most $n$ sample units. In fact, randomly generating some certain weight vectors of length $n$ suffices to approximate the resampling process.

\section{Future study}

To conclude the article, we would like to discuss here a number of interesting topics for future study. First, for datasets with massive sizes, a distributed system is definitely needed. Obviously, there could be no place to store the data. On the other hand, for datasets with sufficiently small sizes, traditional memory based statistical methods can be immediately used. Then, there leaves a big gap between the big and small datasets. Those middle-sized data are often of sizes much larger than the computer memory but smaller than the hard drive. Consequently, they can be comfortably placed on a personal computer, but can hardly be processed by memory as a whole. For those datasets, their sizes are not large enough to justify an expensive distributed system. They are also not small enough to be handled by traditional statistical methods. How to analyze datasets of this size seems to be a topic worth studying. Second, when the whole data are allocated to local machines randomly and evenly, the data on different machines are independent and identically distributed and balanced. Then, all of the methods discussed above can be safely implemented. However, when the data on different machines are collected from (for example) different regions, the homogeneity of the local data would normally be hard to satisfy. The situation could be even worse if the sample sizes allocated to different local machine are very different. How to cope with these heterogeneous and unbalanced local data is a problem of great importance \citep{wang2020efficient}. The idea of meta analysis may be applicable to these situations \citep{liu2015multivariate, zhou2017scalable, xu2020meta}. Finally, in the era of big data, personal privacy is under unprecedented threat. How to protect users' private information during the learning process deserves urgent attention. In this regard, differential privacy (DP) provides a theoretical approach for privacy-preserving data analysis \citep{dwork2008differential}. Some related works associated with distributed learning are \cite{agarwal2018cpsgd, truex2019hybrid,  wang2019differential} and many others. Although it is a hot research areas nowadays, there are still many open challenges. Thus, it is of great interest to study the privacy-preserving distributed statistical learning problem practically and theoretically.

\bibliographystyle{apalike}
\bibliography{ref.bib}

\end{document}